\documentstyle[12pt]{article}
\newcommand{\cent}[1]{\begin{center}#1\end{center}}

\begin{document}
\bibliographystyle{unsr}
\cent{\Large Fragmentation Phase Transitions in Atomic Clusters\\ III\\
--~Coulomb Explosion of Metal Clusters~-- }
\cent{\large O. Schapiro,  P.J. Kuntz, K. M\"ohring, P.A. Hervieux$^*$, \\
D.H.E.~Gross and M.E. Madjet\\
Hahn-Meitner-Institut
Berlin, Bereich Theoretische Physik, \\14109 Berlin, Germany\\
and\\
Freie Universit\"at Berlin, Germany\\~\
$^*$Institut de Physique, Lab LPMC, 1 Bd Arago, F57078 Metz Cedex 3,
France }
\date{\today}
\normalsize

\begin{abstract}
We discuss the role and the treatment of polarization effects in
many-body
systems of charged conducting clusters and apply this to the statistical
fragmentation of Na-clusters. We
see a first order microcanonical phase transition in the fragmentation of
$Na^{Z+}_{70}$ for $Z=$~0 to 8. We can distinguish two fragmentation
phases, namely evaporation of large particles from a large residue and a
complete decay into small fragments only. Charging the cluster shifts
the transition to lower excitation energies and forces the
transition to disappear for charges higher than $Z=8$. At very high charges
the fragmentation phase transition no longer occurs because the
cluster
Coulomb-explodes into small fragments even at excitation energy $\epsilon^* = 
0$. \end{abstract}

\section{Introduction: statistical picture}

Atomic clusters are especially interesting as a link between macroscopic
matter and a single atom. In studying clusters one would usually like to
understand
how the material properties evolve with the size of the system. Our special
interest is slightly different. We would like to address the very
general question of the thermodynamics of finite systems. In this context finite
means that the size of the system is of the same order of
magnitude as
the range of the forces. This imposes an overall correlation within the system
which strongly influences its properties. Atomic clusters together
with nuclei and gravitational systems are the examples of finite systems.

This paper is the third in a series on {\em "Fragmentation Phase Transitions in
Atomic Clusters"}. In the first paper subtitled {\em "Microcanonical
Thermodynamics"} \cite{clust1} we addressed the most general concepts, and in
the second {\em "Symmetry of Fission of Charged Metal Clusters"} \cite{clust2}
we work out how the material properties influence the fragmentation behavior. In
the current work we investigate the influence of charge and polarization effects
on the fragmentation of cold and hot Na-clusters and discuss it in light of
microcanonical fragmentation phase transitions. The next work planned is on the
relation of the fragmentation transition to the commonly known
liquid-gas phase transition.

Our overall goal is to calculate the accessible all-possible particle phase space
$\Omega$ after the fragmentation of a cluster within a given volume. The size of
this accessible phase space is restricted by the conservation of the global
parameters: mass, charge, total energy and the total angular momentum,
which is set to zero in all calculations presented here. In addition there are
geometrical constraints: the fragments within the cluster
configuration should not overlap after the decay
and, moreover, should not be closer than the binding distance of the daughter
clusters. In studying a microcanonical system and applying Microcanonical
Thermodynamics, we need to define the total volume which is accessible to the
fragmenting system. The size of this fixed volume is also one of the
geometrical constraints.

In using the MMMC (Microcanonical Metropolis Monte Carlo) model
\cite{clust1,darius95} we simulate the accessible
phase space at the moment after the fragmentation when the short-range
Van-der-Waals forces (and the exchange of atoms) become negligible. This
happens at the average surface
distances between the fragments of 0.5 to 1~$\AA$. We
simulate the fragmenting configurations within a spherical
volume of radius $R_{sys}=r_f*N_T^{1/3}$, where $N_T$ is the number
of atoms in the
cluster and $r_f=3.85 \AA$ for Na-clusters. $R_{sys}$ is called the {\em
freeze-out} radius.
We want to mention in advance that the results of our calculations are
practically
independent of $R_{sys}$, if $r_f$ is kept within reasonable limits. We place
all the fragments randomly inside this freeze-out volume
under the constraint that  the surfaces are
further apart than a binding distance $d_b\approx 0.5$ to $1\AA$ (see
Appendix~C).

It is also worth noting how the material dependence is included. The ground state
binding energies of all possible fragments make up the input
to our calculation. For sodium we take the known experimental binding energies for
neutral and singly charged clusters up to mass 21 \cite{brechignac89,kappes88}.
Since we
do not know the binding energies of other clusters we determine them from
the liquid drop model. Thus the electronic shell effects are included only for
smaller clusters, which are, however,
the most relevant to the fragmentation. The
weights of individual events (fragmentation configurations) are also dependent
on parameters like the Wigner-Seitz radius, moments
of inertia of the fragments, and the vibrational frequencies of dimers and
trimers.
We treat the internal excitations for all the fragments with $N_j\geq 4$ as
in the bulk matter where we use the bulk specific entropy
(see references \cite{darius95,clust2} for details).

In studying the Coulomb explosion it is important to take
polarization effects into account. We do this by calculating the pairwise
interactions between the fragments, which we assume to be spherical (See
Appendices~A and B). The monomers are considered not to be polarizable.

\section{Coulomb explosion}

The investigation of the stability of charged water droplets goes back to the
work
of Lord Rayleigh \cite{rayleigh} in the year 1882. His ideas were widely applied
to similar problems in nuclear and atomic cluster physics. Afterwards,
scientists in both of these fields were interested especially
in binary fission induced by Coulomb instability as a possible decay channel
\cite{baladron89}. In our calculations we allow {\em all
possible} decay channels
of highly charged clusters. This enables us to study the competition between
evaporation, fission, multifragmentation, or even total vaporization.

The first experimental evidence for Coulomb instability of clusters was
presented
by Sattler et al., \cite{sattler81} for small doubly charged clusters. Some
experimental techniques for achieving highly charged states for masses $N>1000$
are
discussed in \cite{naeher92}. Here we want to concentrate on masses less than
100 in order to study highly charged states by including the polarization effects
and their connection to general questions of thermodynamics in small systems.
A comparison with experimental results is a matter for the future.

Generally, the microcanonical phase transition in a small system is signaled by
an S-shape in the caloric equation of state $T(E)$
\cite{gross95,hueller94a,gross96},
or in other words, a decrease in temperature with growing energy of the system
in the transition region: the system cools while it absorbs energy.
The thermodynamic temperature $T$ is obtained by calculating
the specific entropy, $s$, from a knowledge of
the phase space $\Omega(\epsilon_{tot})$, where $\epsilon_{tot}=E/N$ is the
specific total energy:
\begin{eqnarray}
\beta \equiv \frac{1}{T}& = & \frac{\partial s}{\partial \epsilon_{tot}}
\label{beta} \\
s & = & \frac{ln(\Omega(\epsilon_{tot}))}{N}. ~~~~~~~ \nonumber
\end{eqnarray}
For a fixed mass and charge of the cluster, $\beta$ depends only on the
specific excitation energy $\epsilon^*$,
\begin{eqnarray}
\epsilon_{tot} & = & \epsilon^* + \epsilon_{binding}(N,Z) \nonumber \\
\epsilon_{binding}(N,Z) & = & \frac{ E_{binding}^{N,Z}}{N} ~~.
\end{eqnarray}
$\epsilon_{tot}$ is the total specific energy available to the fragmenting system and
is a globally conserved quantity. $\epsilon^*$ is also a conserved quantity , but
only for fixed $N$ and $Z$. In this case, considering the temperature
dependence, it is sufficient to look at $T(\epsilon^*)$.

We study the fragmentation of clusters at a fixed volume, as we assume that the
motion of complex fragments at short relative distances is dissipative and
ergodic.

In order to concentrate on Coulomb and polarization effects,
we have examined the
statistical fragmentation behavior of the cluster Na$^{Z+}_{70}$ at different
charges Z. Figure~\ref{na702} shows the fragmentation of
Na$_{70}^{2+}$ for a calculation done assuming  pairwise interactions 
between fragments represented by conducting
spheres for a  minimal surface distance
$d_{b}=1\AA$ (See appendices).
The left scale shows the three largest masses M1, M2, M3 of the
fragments (averaged over all configurations) as a function of the specific
excitation energy $\epsilon^*$. We see that
at $\epsilon^*$ below 0.2~eV the system mostly evaporates monomers.
As $\epsilon^*$ increases, some heavier
particles start to appear, but the most dramatic effect is the abrupt
disappearance of the dominating large residue
between 0.35 and 0.45~eV. At even larger excitations the cluster decays only
into small fragments with masses less than 10.
Looking at the thermodynamic temperature $T$,
(the right scale in fig.~\ref{na702}),
we see an S-shape in $T(\epsilon^*)$ around $\epsilon^* \approx $~0.4~eV or
$T\approx$~1200~K. This S-shape is connected with the disappearance
of the large residue and is a typical signature of a microcanonical first
order phase transition.

The average masses M1, M2 and M3 give of course only a rough indication about
the real mass distribution at any given energy.
Figure~\ref{na702m} shows the detailed mass distributions for the same
system as in fig.~\ref{na702} for three individual excitation energies
corresponding to just
before, during, and after the phase transition. The picture does not show the
monomers and does not distinguish between the fragments with the same mass but 
different charge. Before
the phase transition (top panel) Na$_{70}^{2+}$ prefers to evaporate singly
charged trimers
and Na$_{9}^+$. This is due to the shell effects which we include for small
clusters ($N<21$).
For lower excitation energies $0.2<\epsilon^*< 0.25$eV  the $Na^+_3$
channel is much more important than the $Na^+_9$ channel \cite{clust2}. The
predominance of the trimer channel is consistent with molecular-dynamics
predictions
for clusters with $4\le N \le 12$, \cite{Landman91} and with the experimental
findings in \cite{brechignac91}. In our case, however, the occurrence
of $Na^+_3$
cannot be explained just by the least fission barrier, assumed in \cite{brechignac91}. We do not
restrict the number and type of the outcoming fragments; therefore, the choice
of
the system to {\em fission} instead evaporation or multifragmentation in the
energy interval
$0.2<\epsilon^*<0.3$ is not just an energetic, but mainly an entropic effect.

The large residue in figure~\ref{na702m}, top, has preferred masses
Na$_{61}^+$, Na$_{67}^+$ and Na$_{69}^{2+}$,
corresponding to the emission of exactly one fragment with mass N~=~9, 3 or a
neutral monomer.
Inside the phase transition region (middle panel) we see a very broad
distribution of residues combined with two dominating fragments, neutral
Na$_{4}$ and Na$_{9}^+$. The two available charges go mostly to the large
residue and  Na$_{9}^+$; other singly charged small fragments also occur but
with much lower probabilities. One also finds channels with doubly charged
residues with masses 47 to 61, but this decay channel is even less probable.

After the phase transition (lower panel) no large residue is left and the system
decays only into small fragments. The charges prefer to go to two
Na$_{9}^+$ clusters. Again, other singly charged clusters with masses
from 3 to 15 exist, but their probabilities are at least one order of magnitude
lower.

Before studying Coulomb explosion we would like to test the importance of
polarization effects at different charges. Therefore we compare the conducting
spheres (cs) calculation to the point charge (pc) calculation for Na$_{70}^{2+}$
in figure~\ref{na702dt}. A related question is that
of the binding distance $d_b$ between the clusters, because the polarization is
strongly dependent on distances. In Appendix~C we estimate $d_b$ as 0.5 to
1~$\AA$. In figure~\ref{na702dt} we try both $d_b$ values.
We see that for $Z=2$ the thermodynamic temperature, $T(\epsilon^*)$,
is insensitive to $d_b$ and polarization. At the same time one can see that the
statistical uncertainty in $T$ is about $\pm$~70~K for
$\epsilon^*<0.4$~eV. For larger energies the fluctuations become smaller.

Let us now proceed to higher charges. Figure~\ref{na708} is the same as
figure~\ref{na702} but for Na$_{70}^{8+}$. Note the change in
temperature and energy scales. The disappearance of the large residue at
$\epsilon^*\approx$~0.09~eV coincides with a systematic small S-shape
in $T(\epsilon^*)$, indicating
a phase transition. For this charge, the signature of the
phase transition is of the same order of magnitude as the
statistical fluctuations of $T(\epsilon^*)$ in our
calculations. For even higher charges the phase transition disappears. For $Z=8$
the phase transition is shifted significantly towards lower excitation energies
and temperatures for the simple reason that the total energy
which is available to the fragmenting system is increased by the Coulomb
self-energy. Therefore the cluster needs much less additional excitation to
undergo the phase transition. The lower transition temperature $T$ indicates
that by increasing the excitation energy, the additionally gained phase space is
much less here than for Na$_{70}^{2+}$. This is quite natural,
since the phase space
for $Z=8$ is more strongly restricted by the need to distribute all 8
charges among the small fragments.

Looking in figure~\ref{na708m} at the mass distributions for Na$_{70}^{8+}$
before, during and after the phase transition, we see a slightly
different picture from that for Na$_{70}^{2+}$.
We find that at these three energies practically all the fragments,
including the residue, are singly charged.
In the top and middle panels, the small
fragments are predominantly Na$_3^+$ and Na$_9^+$, whereas,
just after the phase
transition (lower panel), they are Na$_7^+$ and Na$_9^+$. The charged trimers
disappear because the
system tries to put all available mass into charged fragments. At even
higher energies the charged trimers appear again together with neutral
fragments.

To illustrate the role of polarization effects, we compare
figure~\ref{na708dt} ($Z=8$) with figure~\ref{na702dt} ($Z=2$).
The statistical fluctuations make the interpretation somewhat difficult,
but two important things are still evident: First, looking at the solid and 
dotted curves for the conducting spheres interactions (cs) we see that the 
phase transition, present for $d_b=1~\AA$, is absent for $d_b=0.5~\AA$. It is
intuitively clear, that there should be some charge $Z$ which makes the cluster
unstable even at $\epsilon^*=0$ and it explodes into small fragments without any
additional excitation. Taking the binding distance $d_b=1~\AA$ this critical
charge is $Z=9$, as we show later. Reducing $d_b$ to $0.5~\AA$ we allow the
fragments
to be placed closer, so that attractive many-body polarization forces become
even more important and cannot be compensated by repulsive
forces any more. This additionally gained energy can be now spent to produce
small fragments. The amount of this energy is in this case already
sufficient to
eliminate the phase transition. (Compare with figure~\ref{na708d}.)
Performing the same calculation for
$d_b=0.5~\AA$ for the point charge interaction (pc) and thereby
neglecting the polarization effects, we still see evidence for a
phase transition (supported also by appearance and later disappearance of the
large residue, as shown in figure~\ref{na708pcd}).
The second observation is that,
outside the transition region ($\epsilon^*<0.04$~eV and $\epsilon^*>0.09$~eV),
the temperature is systematically higher for conducting spheres
than for point charges. For higher
excitations, this shift is larger because of the appearance of neutral
fragments, which are especially sensitive to attraction by polarization. Taking
polarization effects into account reduces the total Coulomb energy. The
system then prefers to put this gained energy into the kinetic energy of the
fragments, (See figure~\ref{energy}, lower part.), increasing the
temperature of the fragmenting configuration.

Figure~\ref{energy} shows the partitioning of the energy into
Coulomb, kinetic, excitation and binding
energy for decaying Na$_{70}^{2+}$ and Na$_{70}^{8+}$.
The energies are normalized with $-E_{tot}$.
We see that the Coulomb energy of the decaying
configuration is negligible for $Z=2$, but becomes absolutely dominant for
$Z=8$. Nevertheless, it is not the Coulomb effects
but rather the change in mass partition that is
significant for the fragmentation phase transition, and
this is of the same type for both high and low charge.

Extending now our results to a whole range of charges, we show the
caloric equation of state $T(\epsilon^*)$ for
charges Z=0...10 in figure~\ref{all_N70}. To help in orientating ourselves
in this
and the following diagram, we indicate some positions with A, B and C. The
dotted
curve (A,C) marks the fragmentation phase transition.
High charges on the cluster
suppress the transition signature and make it disappear for  $Z>8$.

To stress again the significance of the mass partition,
figure~\ref{all_N70m} shows the average mass of the biggest fragment
as a function of $\epsilon^*$ for charges Z=0...10.
The dotted curve (A,C) again indicates
the fragmentation phase transition. Since, independently of the
charge, the transition is signaled by the disappearance of the large residue
we can identify the two phases simply by the presence or absence of
one dominating large residue in the fragmented configuration. Since this is a
microcanonical first order phase transition, there is also a coexistence region
in a certain energy and temperature range (S-shape in $T(\epsilon^*)$).
In this region, the fragmentation events can belong to one or both phases
but outside it, only to one of them.

The fundamental thermodynamic quantity is not the
temperature, but $\beta$. (See equations~\ref{beta}.)
Figure~\ref{beta_N70} shows $\beta(\epsilon_{tot})$ in dependence on the {\em 
total} available energy,
for Na$_{70}^{Z+}$ for charges Z=2...10. We see that keeping the total energy
$\epsilon_{tot}$ constant and increasing the charge reduces the
available kinetic energy and increases $\beta$.
This is a simple way to see how a {\em non-energetic change} of the
system modifies the temperature.

In this paper we studied the N-body phase space of Na$_{70}^{Z+}$-clusters as a
function of charge $Z$ and excitation energy.
We conclude that in the statistical fragmentation process
of metal clusters we can see microcanonical fragmentation phase transitions for
different cluster charges. We can identify the two phases by the
presence or absence of a big residue in the fragment configuration. Very high
charges destroy the phase transition and force the cluster to explode
into small fragments even at an excitation energy $\epsilon^*=0$. For
Na$_{70}^{Z+}$
this happens for $Z > 8$.
The predicted charge and mass distributions in such a
decay process might be also accessible experimentally. 
A question still open is
whether a dynamical fragmentation follows the details of the phase space. This 
is connected to the question as to whether the dynamics is
sufficiently ergodic or, in other words,
whether an experiment is an equilibrated process. Different methods of
exciting or charging the cluster might lead to different results
concerning the equilibration condition.
If this condition is reached, one should be able to see the signature of a phase
transition in a suitable experiment.

\section*{Appendix~A: Polarization effects in many-body sys\-tems and the
approxima\-tion by the two-body inter\-actions}

Coulomb interaction is a long-range interaction which is responsible for the
overall correlation of the fragmenting system. This interaction contributes
significantly to the total (conserved) energy. In our
calculation we consider all the fragments to be conducting and spherical 
because alkali clusters are metallic even for very small sizes 
\cite{brechignac89}. In a
fragmenting many-body configuration the polarization effects are important.
It is known how to treat the polarization of two spherical droplets
from \cite{Krappe}.
Here we address the questions of how to treat polarization effects in a
N-body system and how good the spherical approximation is for very small
clusters.

Generally, calculating Coulomb energy of a conducting sphere with charge $Z$ 
and radius $R$ in presence of a point charge $Z_P$ at distance $d$ from 
its center we need to evaluate 
the interaction of an inhomogenious charge distribution on the 
surface of the sphere with this point charge $Z_P$ outside. 
Evaluating $E_{coul}$ by the method of image charges we substitute this
complicated interaction by the
sum of two point charge interactions: first, $Z_P$ with the charge 
$Z_C$ in the center of the sphere, and second, $Z_P$ with its image 
charge $Z_I$ which is at distance $d_I$ from the center of the sphere 
\cite{jackson}. The Coulomb energy resulting
from the surface charge distribution is exactly equal to the $E_{coul}$ from
the image charge method if
\begin{equation}
Z_I=-\frac{R}{d}Z_P,~~~d_{I}=\frac{R^2}{d};
\end{equation}
\begin{equation}
Z_C=Z-Z_I~~,
\end{equation}
because then the surface of the sphere is an equipotential surface. This is
exactly the condition for the distribution of surface charge on a conducting
sphere.

In our case we want to calculate a Coulomb energy of several charged or neutral
spherical clusters
at close distances. We are going to iterate the image charge method which 
gives us at every iteration step a series of image charges which keep the
surface of each conducting sphere at constant potential. The total interaction
energy is the interaction of image charges of each sphere with all image
charges in other spheres. Therefore the exact
value of $E_{coul}$ can be achieved numerically with any desired accuracy by
performing this iteration.

Let us consider for simplicity first
the case of only two clusters A and B with positive 
charges $Z_A$ and $Z_B$ and
radii $R_A$ and $R_B$ at a distance $d_{AB}$ of their centers, 
see figure~\ref{twosph}. Neglecting any polarization effects in the 0 
iteration step gives the Coulomb energy $E_{0}$ as:
\begin{equation}
E_{0}=\frac{Z_A Z_B}{d_{AB}}.
\end{equation}
In the first iteration step $(1)$ we consider that the point charge $Z_A$ 
induces a negative image charge $Z_{B1}^{(1)}$ on the cluster B, 
and vice versa. The image charges are at
distances $d_{A1}$ and $d_{B1}$ from the centers of clusters A and B and are on
the  line connecting both centers.
\begin{equation}
Z_{A1}^{(1)}=-\frac{R_A}{d_{AB}}Z_B,~~~d_{A1}=\frac{R_A^2}{d_{AB}};~~~~~~~~~
Z_{B1}^{(1)}=-\frac{R_B}{d_{AB}}Z_A,~~~d_{B1}=\frac{R_B^2}{d_{AB}}.
\end{equation}
To account for charge conservation one needs to subtract the charge 
$Z_{B1}^{(1)}$ in the
center of cluster B and $Z_{A1}^{(1)}$ in the center of cluster A.
We get for the central charge:
\begin{equation}
Z_{A0}^{(1)}=Z_A-Z_{A1}^{(1)};~~~~~~~~~
Z_{B0}^{(1)}=Z_B-Z_{B1}^{(1)}.
\end{equation}
The Coulomb energy $E_{1}$ after the first iteration step
is then:
\begin{equation}
E_{1}=
\frac{Z_{A0}^{(1)}Z_{B1}^{(1)} }{ d_{AB}-d_{B1}}   +
\frac{Z_{B0}^{(1)}Z_{A1}^{(1)} }{ d_{AB}-d_{A1}} +
\frac{Z_{A1}^{(1)}Z_{B1}^{(1)} }{ d_{AB}-d_{A1}-d_{B1}} +
\frac{Z_{A0}^{(1)}Z_{B0}^{(1)} }{ d_{AB}}.
\end{equation}

In the second iteration step we need first to adjust the image charges at 
positions $d_{A1}$ and $d_{B1}$, since the charges in the centers were changed.
\begin{equation}
Z_{A1}^{(2)}=-\frac{R_A}{d_{AB}}Z_{B0}^{(1)};~~~~~~~~~
Z_{B1}^{(2)}=-\frac{R_B}{d_{AB}}Z_{A0}^{(1)}.
\end{equation}

Further, the image charges $Z_{A1}^{(2)}$ and $Z_{B1}^{(2)}$ induce the second
order image charges $Z_{A2}^{(2)}$ and $Z_{B2}^{(2)}$.
\begin{equation}
Z_{A2}^{(2)}=-\frac{R_A}{d_{AB}-d_{B1}}Z_{B1}^{(2)},
        ~~~d_{A2}=\frac{R_A^2}{d_{AB}-d_{B1}};
\end{equation}
\begin{equation}
Z_{B2}^{(2)}=-\frac{R_B}{d_{AB}-d_{A1}}Z_{A1}^{(2)},
        ~~~d_{B2}=\frac{R_B^2}{d_{AB}-d_{A1}}.
\end{equation}
For the Coulomb energy we get using $d_{A0}=d_{B0}=0$,
\begin{equation}
E_{2}=
\sum_{i,j=0}^2 {
\frac{Z_{Ai}^{(2)}Z_{Bj}^{(2)} }{ d_{AB}-d_{Ai}-d_{Bj}}
}.
\end{equation}

In every next iteration step $n$ we consider two aspects: 1) The charges in the
center have been changed, therefore we need to adjust the image charges 
$Z_{A1}^{(n)}$
and $Z_{B1}^{(n)}$, and 2) these image charges 
induce themselves second-order image
charges $Z_{A2}^{(n)}$ and $Z_{B2}^{(n)}$, and so on up to  
$Z_{An}^{(n)}$ and $Z_{Bn}^{(n)}$. For the Coulomb energy we need
to account for contributions from all the image charges in sphere A with all
image charges in sphere B:
\begin{equation}
\label{energy1}
E_{n}=
\sum_{i,j=0}^n {
\frac{Z_{Ai}^{(n)}Z_{Bj}^{(n)} }{ d_{AB}-d_{Ai}-d_{Bj}}
}.
\end{equation}

We can show \cite{thesis} that $E_n$ build a convergent sequence of energies.
If charges $Z_{A}$ and $Z_B$ have the same sign then 
\begin{equation}
\label{series}
E_{0} > E_{2} > E_{4} > ~...~E_{n}~...~> E_{3} > E_{1}~~.
\end{equation}
Please note that the even subscripts are on the left side 
the odd ones on the right.

The extension of the method to 3 or any number of conducting spheres is only a
bookkeeping problem. 
Figure~\ref{threesph} shows, for example, the positions of image charges in the
second iterations step for the case of three clusters. 
Let us concentrate on the left picture 
showing the complete collection
of image charges. Besides the image charges on lines connecting the centers,
which are represented by circles, we
get image charges resulting from the interaction of the first-order images in
one sphere with the other two spheres. The positions of these images are 
shown by triangles. Calculating energy we use equation~\ref{energy1} extending
it to the general case by the summation over all the spheres.

It is also clear that this method is valid if some of the
conducting spheres are neutral. They are attracted due to polarization by the
charged clusters.

Our main aim is to find a representative set of configurations in phase space
$\Omega$. For 1000000 configurations with 3 to 4 fragments it is no big problem
to calculate the Coulomb interaction exactly. But for most of the excitation
energies we need configurations with many more fragments.
The exact treatment of the Coulomb energy makes the statistical task
unsolvable within a reasonable computation time. Therefore we investigated
whether we can separate the many-body interaction into the sum of
two-body interactions.

We first concentrate on a three-body case, see again figure~\ref{threesph}. 
Considering only the two-body interactions, right picture, we neglect the image
charges symbolized by triangles in the left picture.

We describe the positioning of three
clusters by two variables, the distance $d_C$ of the center of the 
sphere C to the center of mass of A and B and $d_{AB}$ the distance
between the centers of A and B.
Figure~\ref{3_body_e} shows the Coulomb energy surface for three Na$_{10}^+$
clusters.
The plateau at the small $d_{AB}$ and $d_{C}$ is a forbidden area where the
fragments would overlap. Figure~\ref{3_body_de} is the difference $\Delta E$
between
the exact three-body interaction and the sum of the two-body interactions. The
difference is smaller than 2\%.  Testing different combinations of masses
$N=10$ and $N=100$ of the
three singly or multiply charged fragments always yields $\Delta E < 5$\%. The
same is true for one or two small neutral clusters in the presence of
charged ones. This is well within the desired accuracy. 

Generally, the
polarization part of the total Coulomb energy increases for larger clusters and
smaller charges. The polarization part becomes especially dominant for a big
neutral cluster in the presence of small charged ones at {\em very} short
distances. In this case the approximation by a sum of
two-body interactions gives an arbitrarily large deviation from the exact
many-body calculation. In our fragmentation calculation of a charged metallic
cluster we {\em never} get configurations where the whole charge goes into small
fragments aside one big neutral residue if the fragments have to be placed at
surface distances larger than a binding distance $d_b \approx 0.5 \AA$.
Therefore we use the two-body conducting sphere approximation,
evaluating the Coulomb energy in the MMMC
calculation of metallic cluster fragmentation.

Finally, to illustrate the role of polarization effects we have shown a
calculation for the fragmentation of Na$_{70}^{2+}$ and Na$_{70}^{8+}$ in
figures \ref{na702dt} and \ref{na708dt}. We get
systematically lower temperatures for the point charge interaction in
comparison to the pairwise conducting
spheres interaction. This means that
taking the point charge interaction makes a smaller phase
space accessible than in the case of conducting spheres.

\section*{Appendix~B: Polarization effects for dimers and trimers}

Most of the fragmentation configurations have many small fragments such as
dimers and trimers. It is not obvious that treating dimers and
trimers as spheres for calculating polarization effects is a good
approximation.
In order to clarify this, we calculated the interaction of the $\rm Na_3^+$
trimer-ion with a charged conducting sphere by means of the
diatomics-in-molecules (DIM) method, modified in such a way as to take
into account the electrostatic interaction between the ion and the sphere as
well as the mutual polarizabilities\cite{Kuntz}.
The DIM method yields the potential
energy surface of the trimer as a function of the inter-nuclear distances
and can correctly describe the dissociation of the trimer-ion into
$\rm Na_2^+ +Na$ or $\rm Na_2 +Na^+$.  In the presence of the sphere,
the charge distribution in the molecule and the geometrical configuration
may alter because of the electrostatic interactions.  We found that
the energy as a function of distance from the sphere is nearly the
same as that calculated from the electrostatic energy alone provided that the
charge distribution and geometrical configuration are allowed to relax
freely.  The energy also depends on the orientation of the trimer, there
being two local minima at a fixed distance from the sphere: orientation A,
an isosceles triangle having its apex between the sphere and the base,
and orientation B, where the apex lies on the far side of the base as
seen from the sphere. The energies for these two local minima bracket the
electrostatic energy as calculated by treating the trimer as a sphere.
Finally, when the trimer approaches too closely to the sphere, it
becomes unstable because all of the charge flows onto the nearest atom,
which is no longer bound to the neutral dimer.

Figure~\ref{kuntz} shows the two orientations A and B of a trimer corresponding
to minima
in the interaction energies. The curves show the dependence of the
Coulomb interaction energy on the
distance of the center of mass of the trimer to the surface of the cluster of
radius 27 bohr and charge 5. The minimal orientation A is in nice overall
agreement
with molecular dynamics calculation \cite{Landman91}. The dot-dashed
curve corresponds to the interaction between two
conducting spheres, where the radius of the trimer-sphere was again
taken as $r_{trimer}=r_s3^{1/3}$. 
To put this approximation in relation
to more simplified solutions, we show figure~\ref{kuntz1} which is a blow-up of
figure~\ref{kuntz}. The dotted curve is the approximation by point charge
interaction for the charge localized in the centers, 
and the short dashed curve
is when considering the large cluster as a polarizable sphere and 
the trimer as a point charge outside. 

This shows the validity of
a spherical approximation for calculating the polarization 
energies for trimers.

For a dimer the situation is even better, since there is only one minimum in
energy for the orientation radial to the cluster. In this case all the polarized
charge moves as in the image charge method above.

Of course, considering many-body configurations, the dimers and trimers cannot
have the same orientation to all the fragments.
Even though we cannot account for the individual
orientations because of the demands on computer time, we
know at least that the spherical approximation, which is like averaging over all
orientations, gives a reasonable result.

\section*{Appendix~C: The minimal surface distances $d_b$ between the clusters}

Another much more important point is the minimal surface distance
$d_b$ for the clusters not to be bound to each other. It would be natural to
assume that this distance is of the order of magnitude of the
sum of the electronic spill-outs \cite{yannouleas93,yannouleas95} of the two
interacting clusters. The problem is that spill-out is in general charge
dependent and is probably strongly deformed by the presence of another
charge outside the considered cluster.
Therefore we obtained a rough estimate from the DIM
calculation (see Figure~\ref{kuntz}),
which also gives the forces experienced by the three
trimer atoms. The calculation break down when the attraction of
one of the trimer atoms to the cluster is larger than its binding within the
trimer. This minimal distance gives an estimate for the lower limit of $d_b$ as
$d_b \approx 0.5$ to $1\AA$.

\section*{Acknowledgements}

We thank the Sonderforschungsbereich SFB 337 of the Deutsche
Forschungsgemeinschaft for substantial support and the Freie Universit\"at
Berlin for offering us their computation facilities.

\begin{figure}[p]
\caption{\label{na702}Na$_{70}^{2+}$: Average three largest masses M1, M2, M3
(left scale) and the thermodynamic temperature (right scale) as a function of
specific excitation energy $\epsilon^*$. $d_{b}=1\AA$, pairwise conducting spheres
interaction.}
\end{figure}

\begin{figure}[p]
\caption{\label{na702m}Na$_{70}^{2+}$: Mass distributions for different
specific excitation energies $\epsilon^*$}
\end{figure}

\begin{figure}[p]
\caption{\label{na702dt}Na$_{70}^{2+}$: Thermodynamic temperature $T(\epsilon^*)$ for
different $d_b$ values and conducting spheres (cs) versus point charge (pc)
Coulomb interactions.
}
\end{figure}

\begin{figure}[p]
\caption{\label{na708}Like figure
\protect \ref{na702} \protect
for Na$_{70}^{8+}$. }
\end{figure}

\begin{figure}[p]
\caption{\label{na708m}Like figure
\protect \ref{na702m}\protect
for Na$_{70}^{8+}$. }
\end{figure}

\begin{figure}[p]
\caption{\label{na708dt}Like figure
\protect\ref{na702dt} \protect
for Na$_{70}^{8+}$. }
\end{figure}

\begin{figure}[p]
\caption{\label{na708d}Like figure
\protect\ref{na708} \protect
for $d_b=0.5\AA$.}
\end{figure}

\begin{figure}[p]
\caption{\label{na708pcd}Like figure
\protect\ref{na708d}\protect
for point charge interactions.}
\end{figure}

\begin{figure}[p]
\caption{\label{energy} Partition of the total available energy for fragmenting
Na$_{70}^{2+}$ and Na$_{70}^{8+}$.} \end{figure}

\begin{figure}[p]
\caption{\label{all_N70}Na$_{70}^{Z+}$: Caloric equation of state $T(\epsilon^*)$ for
charges Z=0...10. The dotted curve indicates the fragmentation phase
transition.} \end{figure}

\begin{figure}[p]
\caption{\label{all_N70m}Na$_{70}^{Z+}$: Average mass of the biggest fragment in
dependence on $\epsilon^*$ for charges Z=0...10. The dotted curve indicates the
fragmentation phase transition. The Points A, B and C indicate the
corresponding places in figure~
\protect\ref{all_N70}.\protect
} \end{figure}

\begin{figure}[p]
\caption{\label{beta_N70}Na$_{70}^{Z+}$: $\beta(\epsilon^*)$ for
charges Z=2...10. } \end{figure}

\begin{figure}[p]
\caption{\label{twosph} First iteration steps for evaluation of Coulomb interaction energy by the method of image charges in case of two conducting spheres A and B with charges $Z_A$ and $Z_B$.
} \end{figure}

\begin{figure}[p]
\caption{\label{threesph} The second iteration step for evaluation of Coulomb interaction energy by the method of image charges for three conducting spheres A, B and C with charges $Z_{A}, Z_B$ and $Z_C$.
} \end{figure}

\begin{figure}[p]
\caption{\label{3_body_e} Exact three-body Coulomb energy $E$ of three spherical
conducting clusters with mass N=10 and charge Z=1 as a function of different
relative positions, see text, Appendix~A.} \end{figure}

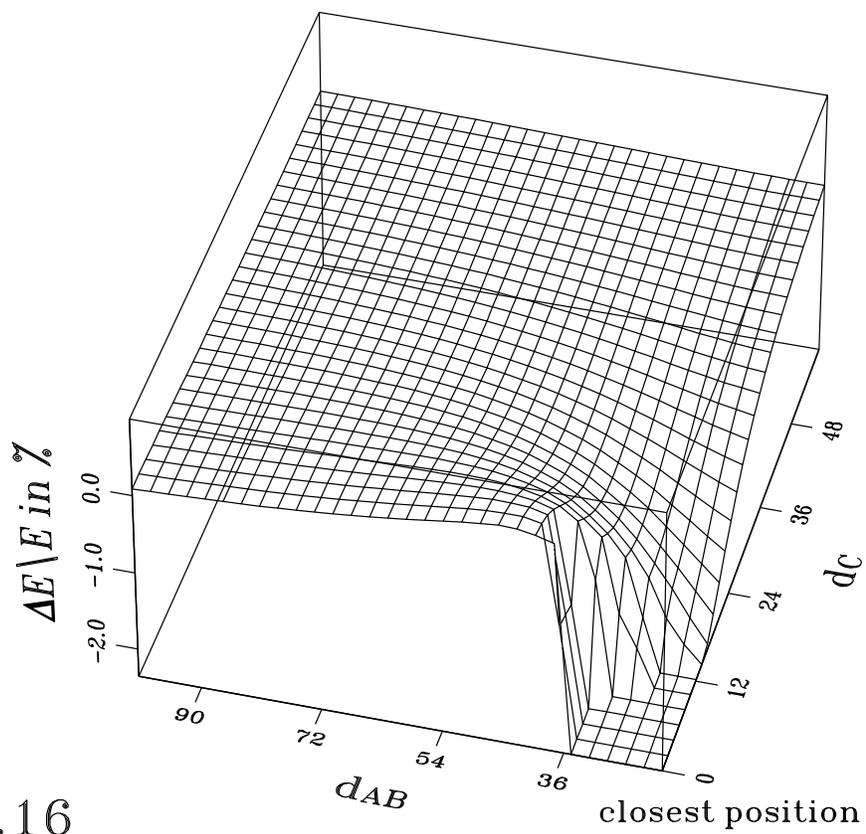
\begin{figure}[p]
\caption{\label{3_body_de} Difference $\Delta E$ between  $E$,
figure~
\ref{3_body_e}
 and the sum of the two-body interactions of three spherical
conducting clusters with mass N=10 and charge Z=1 as a function of different
relative positions, see text, Appendix~A.} \end{figure}

\begin{figure}[p]
\caption{\label{kuntz}Coulomb energy for a charged sodium trimer in a Coulomb
field of a spherical cluster with radius 27 bohr, Z=5 for two orientations A
and B, see text, Appendix~B. The dot-dashed curve is a spherical approximation
of a trimer. } \end{figure}

\begin{figure}[p]
\caption{\label{kuntz1} The three curves from figure~
\protect\ref{kuntz}\protect
are shown together with two other approximations for Coulomb energy: 
Dotted curve considers both, a sphere and a trimer 
having fixed point charges in their centers. Dashed curve shows the 
point charge approximation just for the trimer in presence of  conducting 
sphere.
} \end{figure}

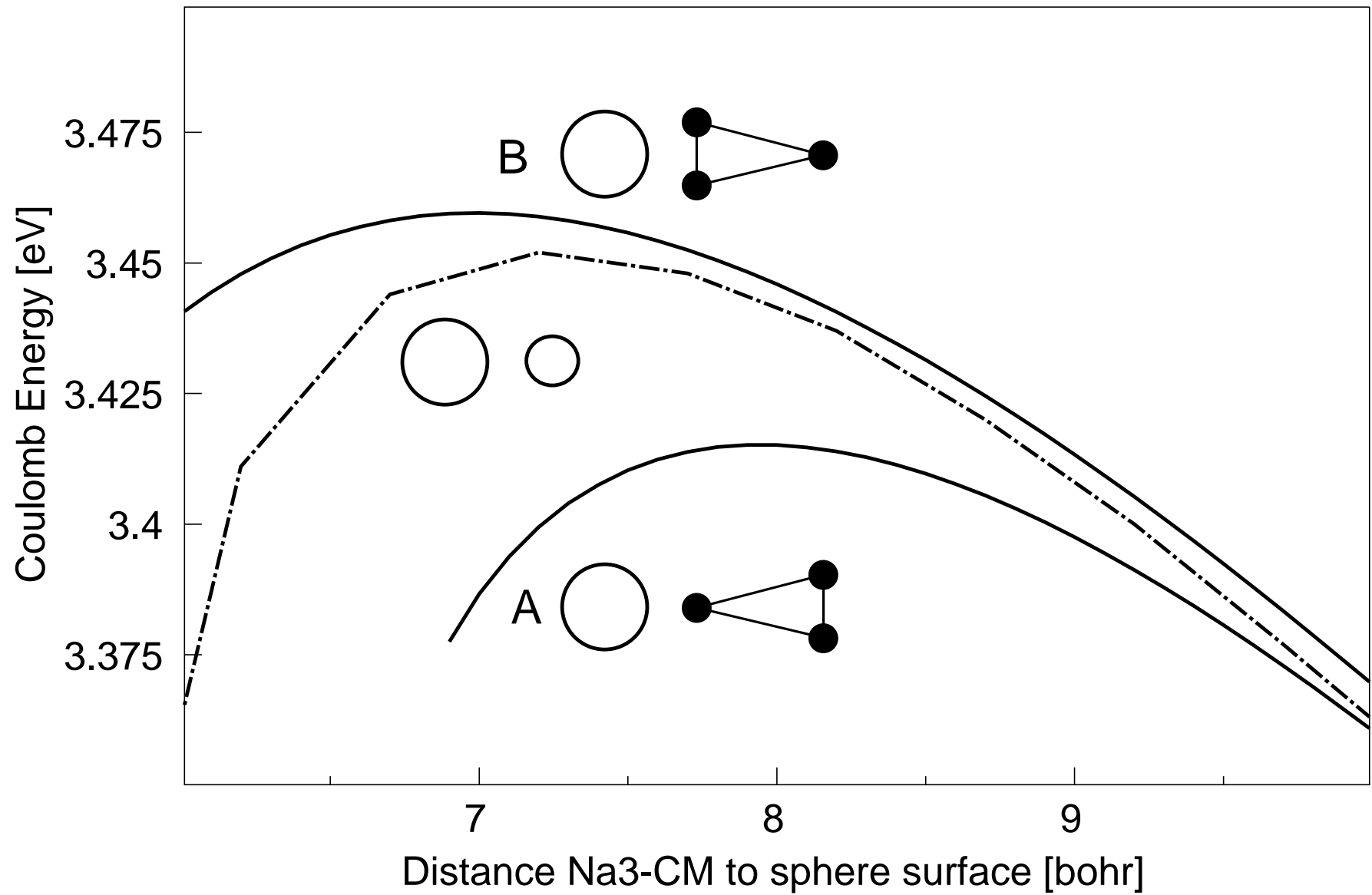
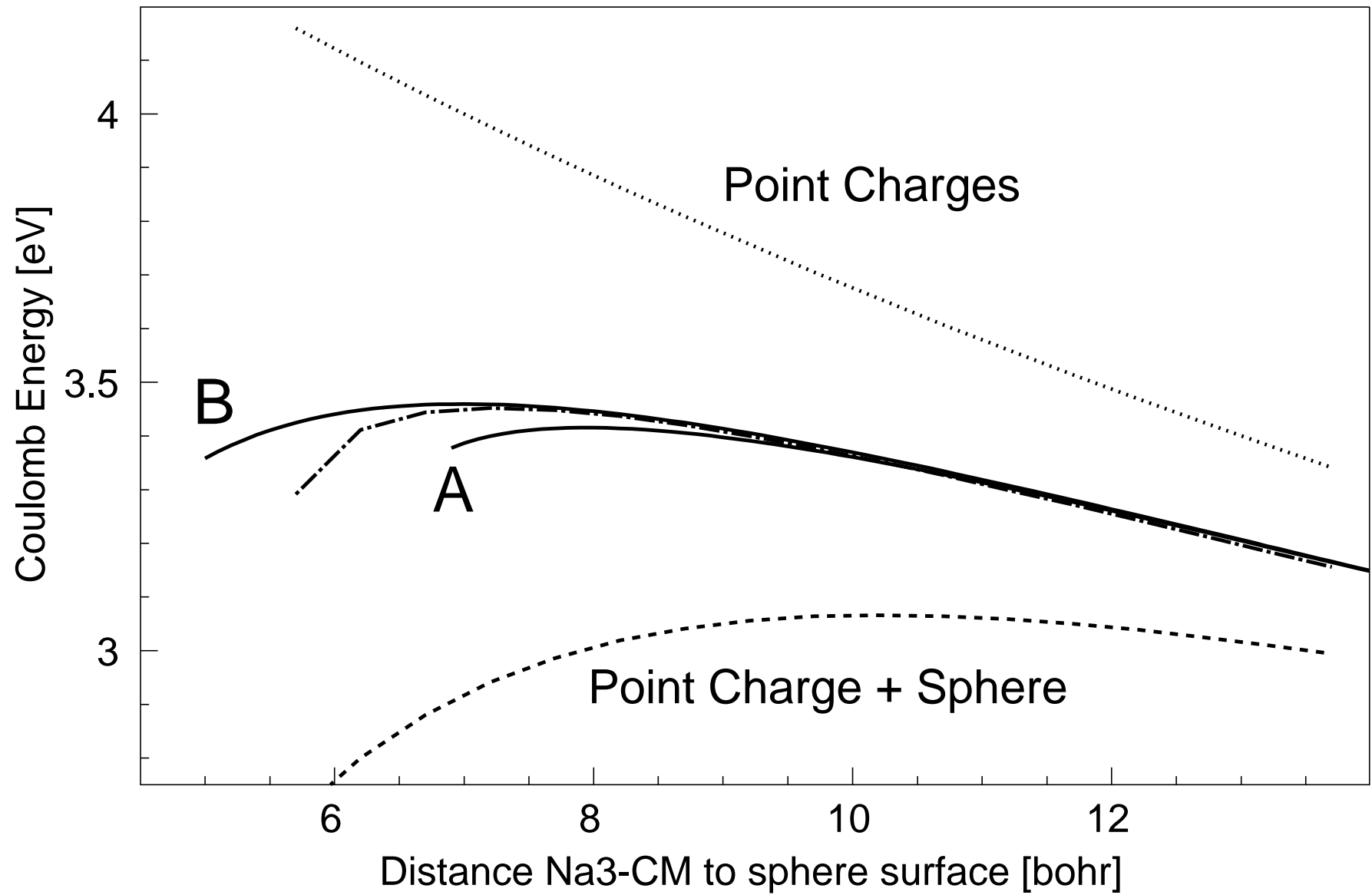
\end{document}